\documentclass[alpha-refs]{wiley-article}

\usepackage{color}
\usepackage{ulem}

\usepackage{siunitx}
\usepackage{amsmath}
\usepackage{graphicx}
\usepackage{booktabs}
\usepackage[colorlinks=true, allcolors=black]{hyperref}
\usepackage{mathtools}  
\usepackage{comment}
\usepackage{multirow}

\papertype{Original Article}
\paperfield{Journal Section}

\title{Flux gradient relations and their dependence on turbulence anisotropy}

\author[1]{Samuele Mosso}
\author[2\authfn{2}]{Marc Calaf}
\author[1\authfn{1}]{Ivana Stiperski}

\affil[1]{Department of Atmospheric and Cryospheric Sciences, University of Innsbruck, Innsbruck 6020, Austria, }
\affil[2]{Department of Mechanical Engineering, University of Utah, Salt Lake City, Utah 84112, USA}

\corraddress{Mosso S., Department of Atmospheric and Cryospheric Sciences, University of Innsbruck, Innsbruck 6020, Austria}
\corremail{samuele.mosso@uibk.ac.at}

\presentadd[\authfn{1}]{ivana.stiperski@uibk.ac.at}
\presentadd[\authfn{2}]{marc.calaf@utah.edu}

\fundinginfo{These results are part of a project that has received funding from the European Research Council (ERC) under the European Union’s Horizon 2020 research and innovation program (Grant agreement No. 101001691)}

\runningauthor{Mosso et al. 2023}

\begin{document}

\maketitle

\begin{abstract} 
Monin-Obukhov similarity theory (MOST) is used in virtually every Earth System Model (ESM) to parameterize the near-surface turbulent exchanges and mean variables' profiles. Despite its wide spread use, there is high uncertainty in the literature about the appropriate parameterizations to be used. In addition, MOST has limitations in very stable and unstable regimes, over heterogeneous terrain and complex orography, and has been found to incorrectly represent the surface fluxes. A new approach including turbulence anisotropy as a non-dimensional scaling parameter has recently been developed, and has been shown to overcome these limitations and generalize the flux-variance relations to complex terrain.

In this paper we analyze the flux-gradient relations for five well known datasets, ranging from flat and homogeneous to slightly complex terrain. The scaling relations show substantial scatter and highlight the uncertainty in the choice of parameterization even over canonical conditions. We show that by including information on turbulence anisotropy as an additional scaling parameter, the original scatter becomes well bounded and new formulations can be developed, that drastically improve the accuracy of the flux-gradient relations for wind shear ($\phi_M$) in unstable conditions, and for temperature gradient ($\phi_H$) both in unstable and stable regime. This analysis shows that both $\phi_M$ and $\phi_H$ are strongly dependent on turbulence anisotropy and allows to finally settle the extensively discussed free convection regime for $\phi_M$, which clearly exhibits a $-{1/3}$ power law when anisotropy is accounted for.

Furthermore we show that the eddy diffusivities for momentum and heat and the turbulent Prandtl number are strongly dependent on anisotropy and that the latter goes to zero in the free convection limit.

These results highlight the necessity to include anisotropy in the study of near surface atmospheric turbulence and lead the way for theoretically more robust simulations of the boundary layer over complex terrain.

\end{abstract}

\section{Introduction}
\label{sec:intro}

The atmospheric boundary layer (ABL) is the region of the atmosphere contained between the Earth's surface and the mostly inviscid free troposphere. As a result, the ABL is driven by the local surface forcing at the bottom (e.g. friction, heating/cooling, evaporation, etc.), and the larger scale flow dynamics of the free atmosphere at the top. It is within this region that most common weather patterns develop, and thus its understanding and accurate representation in Earth System Models (ESMs) of different resolution scales is critical. The flow in the ABL is generally turbulent, with turbulent motions ranging from scales of millimeters (i.e. Kolmogorov microscale) up to a few kilometers (i.e. the height of the ABL, $z_i$), making its full computational representation a chimera with current computing power \citep{calaf2023boundary}. Near the surface, within the so-called atmospheric surface layer (ASL) where the energetic turbulent eddies scale with the distance from the wall, the numerical representation of the physical processes is even more challenging. To bypass these limitations, surface-atmosphere interactions are commonly parameterized using Monin-Obukhov similarity theory (MOST) \citep{monin54,foken200650}. This approach is employed in most ESMs, regardless of whether they are turbulence resolving (e.g. Large-Eddy Simulations, LES), or not (i.e. mesoscale, weather prediction and climate models) and is thus a key component in the correct representation of surface-atmosphere exchange on all scales of motion \citep{Edwardsetal2020}. Here we continue to extend the applicability of MOST from flat and horizontally homogeneous terrain (canonical conditions), for which it was developed, to more general surface conditions characterized by orography and surface heterogeneity \citep[cf.,][]{stiperski2023generalizing}.

MOST states that in the absence of subsidence (mean downward vertical flow), under steady-state conditions (weak time dependence of turbulence statistics), and horizontal homogeneity of the mean flow statistics, turbulence in ASL (characterized by semi-constancy of fluxes with height) can be fully described by the ratio of buoyancy production/destruction and shear production, represented by the non-dimensional length scale $\zeta = z/L$, also known as the stability parameter \citep{stull1988introduction}. 
In this relation, $z$ is the height above ground, representative of the characteristic length-scale of near-surface turbulence, and $L$ is the Obukhov length \citep{obukhov1971turbulence,stull1988introduction} formed by the surface friction velocity ($u_* = ({\overline{u'w'}^2 + \overline{v'w'}^2})^{1/4}$, where $\overline{u'w'}$ and $\overline{v'w'}$ are the streamwise and spanwise momentum flux), the surface buoyancy flux ($\overline{w'\theta_v'}$), buoyancy parameter $(g/\overline{\theta_v})$, and von K{\'a}rm{\'a}n constant ($\kappa \approx 0.4$), such that $L = - {u_*^3\overline{\theta_v}}/{\kappa\,g\overline{w'\theta_v'}}$. According to MOST, any properly non-dimensionalized mean variable in the ASL is solely a function of $\zeta$ \citep{monin54}.

When MOST is applied to the profiles of mean wind and temperature, it leads to the so-called flux-gradient (or flux-profile) scaling relations, defined as

\begin{equation}
\label{eq:grad}
\begin{aligned}
\phi_{M}\left(\zeta\right) & =\frac{\kappa z}{u_*}\frac{d\overline{u}}{dz}, \\ 
\phi_{H}\left(\zeta\right) &=\frac{\kappa z}{\theta_v*}\frac{d\overline{\theta}}{dz}, 
\end{aligned}
\end{equation}
where the temperature scale is defined as $\theta_* = -\overline{w'\theta_v'}/u_*$. In neutrally stratified ASL conditions, i.e. in the absence of background stratification when the buoyancy flux $\overline{w' \theta_v'}$ tends to zero, $\zeta$ also tends to zero, the flux-gradient relations reduce to the famous law-of-the-wall \citep{pope2000turbulent} for momentum $\phi_M \sim 1 $, and similarly for temperature gradient to $\phi_H \sim Pr_t$, where $Pr_t$ is the turbulent Prandtl number \citep{pope2000turbulent}.

In non-neutral conditions the shape of the $\phi_M$ and $\phi_H$ functions requires empirical determination, and was thus initially developed through extensive experimental measurements. For example, the pioneering Kansas 1968 experiment performed over the ideally spatially homogeneous wheat fields, is the basis for the development of the so-called Businger-Dyer flux-gradient relations \citep{businger71}. These relations were modified \citep{hogstrom88}, and ultimately formalized \citep[][hereon referred to as HÖ96]{hogstrom96} to the following form:

\begin{align}
\label{eq:busi} 
\phi_M &= 
\begin{cases}
    (1-\gamma_M\zeta)^{-\frac{1}{4}} & \zeta<0\\
    1+\beta_M\zeta & \zeta >0
\end{cases} 
\qquad \text{and} \nonumber\\[2ex] 
\phi_H &= 
\begin{cases}
    Pr_t\left(1-\gamma_H\zeta\right)^{-\frac{1}{2}} & \zeta <0\\
    1+\beta_H\zeta & \zeta >0
\end{cases},
\end{align}
where the $\beta_M$, $\gamma_M$, $\beta_H$, and $\gamma_H$ are empirical constants given in Tables~\ref{tab:unstablescaling} and \ref{tab:stablescaling}.

Subsequent measurement campaigns allowed the community to derive refined scaling relations that take into account the behavior of $\phi_{M}$ and $\phi_{H}$ in both the very stable and unstable regimes which were understudied in the previous campaigns. In the unstable regime (Table~\ref{tab:unstablescaling} and Figure~\ref{fig:scaling}), where $\zeta<0$, the HÖ96 parametrizations follow $\phi_M \sim(-\zeta)^{-1/4}$ and $\phi_H \sim(-\zeta)^{-1/2}$ power laws. These power laws, however, do not conform to the free convective limit in which $u_*$ is no longer a relevant scaling velocity. Instead, \cite{grachev2000convective} and \citep{li2021keyps} used scaling arguments and analytical theory to predict a $\phi_{M,H} \sim-\zeta^{-1/3}$. Finally, \cite{kader90} proposed a different model based on directional dimensional analysis that postulates the presence of three sub-layers (dynamic, dynamic-convective and convective) with different dynamics in the range of values of $\zeta$. In particular, their model predicts that $\phi_M \sim \zeta^{1/3}$ in the convective sub-layer when $\zeta < -1$, in opposition to the previous arguments. Two equivalent analytical forms of the model are reported in this study by \cite{kader1989effect} and \cite{brutsaert1992stability}.

\belowrulesep = 10pt
\begin{table}
\caption{Parametrizations of the flux-gradient relations in unstable regime.}
\centering
\vspace*{5mm}
\resizebox{\textwidth}{!}{
\setlength{\tabcolsep}{15pt}
\begin{tabular}{lcc}
\toprule
    \large\textbf{Author and abbreviation} & \Large$\Phi_M$ & \Large$\Phi_H$ 
    \\
\midrule
    \begin{tabular}{@{}l@{}}\cite{hogstrom96},\\ \textbf{HÖ96} \end{tabular}    &
    $(1-19\zeta)^{-\frac{1}{4}}$ &
    $0.96\left(1-11.6\zeta\right)^{-\frac{1}{2}}$
    \\
    [15pt]\begin{tabular}{@{}l@{}}\cite{grachev2000convective},\\ \textbf{GR00} \end{tabular}  &
    $\left(1-10\zeta\right)^{-\frac{1}{3}}$ &
    $\left(1-34\zeta\right)^{-\frac{1}{3}}$
    \\ [15pt]
    \begin{tabular}{@{}l@{}}\cite{kader90},\\ \textbf{KY90} \end{tabular}  &
    $\left(\dfrac{1+0.6\zeta^2}{1-7.5\zeta}\right)^{\frac{1}{3}}$ &
    $0.64\left(\dfrac{3-2.5\zeta}{1-10\zeta+50\zeta^2}\right)^{\frac{1}{3}}$
    \\[15pt]
    \begin{tabular}{@{}l@{}}\cite{brutsaert1992stability},\\ \textit{alternative formulation of KY90} \end{tabular}& 
    $\dfrac{0.37 - 0.24 |\zeta|^{0.72}}{0.37 + |\zeta|^{0.72}} - 0.5 \zeta^{1/3}$ &
    $\dfrac{0.33 + 0.057 |\zeta|^{0.78}}{0.33 + |\zeta|^{0.78}}$
    \\[15pt]
\bottomrule
\end{tabular}
}
\label{tab:unstablescaling}
\end{table}

In the very stable regime ($\zeta > 1$), a flattening of $\phi_M $ and $ \phi_H$ is typically encountered, deviating from the linear Businger-Dyer relations \citep[][see Table~\ref{tab:stablescaling} and Fig.~\ref{fig:scaling}]{cheng2005flux,gryanik2020new,beljaars1991flux}. However, \cite{grachev2013critical}, and later \cite{babic2016evaluation}, clearly showed that data with a Richardson number lower than the critical value ($Ri \leq Ri_c \approx 0.21, 0.25$) follow the Businger-Dyer relations, and postulated that the above mentioned flattening stems from the inclusion of non-Kolmogorov turbulence for which MOST is not valid. In this very stable regime, \cite{boykostochastic} used a stochastic approach to derive a two-dimensional probability density function for $\phi_M$, incorporating stochastic effects of submesoscale motions and intermittency into its description.

\begin{table}
\caption{Parametrizations of the flux-gradient relations in stable regime.}
\centering
\vspace*{5mm}
\resizebox{\textwidth}{!}{
\setlength{\tabcolsep}{6pt}
\begin{tabular}{lcc}
\toprule
    \large\textbf{Author and abbreviation} & \Large$\Phi_M$ & \Large$\Phi_H$ 
    \\
\midrule
    \begin{tabular}{@{}l@{}}\cite{hogstrom96},\\ \textbf{HÖ96}\end{tabular} &
    $1+5.3\zeta$ &
    $1+8\zeta$
    \\[15pt]
    \begin{tabular}{@{}l@{}}\cite{cheng2005flux},\\ \textbf{CB05}\end{tabular}   &
    $\begin{multlined}1+6.1 \cdot\\[-2ex]
    \left(\dfrac{\zeta+\zeta^{2.5}(1+\zeta^{2.5})^{\frac{-1.5}{2.5}}}
    {\zeta+(1+\zeta^{2.5})^{\frac{1}{2.5}}}\right)\end{multlined}$ &
    $\begin{multlined}1+5.3\cdot\\[-2ex]
    \left(\dfrac{\zeta+\zeta^{1.1}(1+\zeta^{1.1})^{\frac{-0.1}{1.1}}}
    {\zeta+(1+\zeta^{1.1})^{\frac{1}{1.1}}}\right)\end{multlined}$
    \\[30pt]
    \begin{tabular}{@{}l@{}}\cite{beljaars1991flux},\\ \textbf{BH91}\end{tabular} &
    $\begin{multlined}
        1+\zeta+\\[-2ex]
        \frac{2}{3}\zeta\cdot(6-0.35\zeta)\cdot\\[1ex]
        e^{-0.35\zeta}
    \end{multlined}$ & 
    $\begin{multlined}
        1+\zeta\cdot\left[1 + \frac{2}{3}\zeta\right]^{\frac{1}{2}} +\\[-2ex]
       \frac{2}{3}\zeta\cdot(6-0.35\zeta)\cdot \\[1ex]
        e^{-0.35\zeta}
    \end{multlined}$
    \\[40pt]
    \begin{tabular}{@{}l@{}}\cite{gryanik2020new},\\ \textbf{GR20}\end{tabular} &
    $1+\dfrac{5\zeta}{(1+0.3\zeta)^{\frac{2}{3}}}$ & 
    $0.98\left(1+\dfrac{5\zeta}{1+0.4\zeta}\right)$
    \\[15pt]
\bottomrule
\end{tabular}
}
\label{tab:stablescaling}
\end{table}

The non-uniqueness of scaling relations for $\phi_M$ and $\phi_H$ in the very stable and unstable regimes even over flat and homogeneous terrain points to MOST's limited applicability not only under restrictive surface conditions, but also outside of a narrow stability range. Additionally, in very stable stratification MOST is considered to be affected by self-correlation \citep{klipp04}.
Alternatively, local similarity scaling \citep{nieuwstadt1984turbulent} can be used in cases where the constant flux assumption fails, e.g. stable stratification \citep[e.g.,][]{grachev05} or over complex terrain \citep[e.g.,][]{deFranceschi2009,babic2016evaluation,sfyri18,stiperski19}. In this approach, the local Obukhov length $\Lambda$ that varies with height is used instead of $L$.

Over complex terrain (such as heterogeneous surfaces or orography) multiple assumptions of MOST are clearly violated, and thus the validity of MOST is far from guaranteed. Still, a number of recent studies have shown that if the upstream fetch is homogeneous, MOST can still be valid even over complex terrain, such as for specific wind sectors in an alpine valley \citep{deFranceschi2009} and an arctic fjord \citep{kral2014observations}. At the same time, \cite{babicN2016} observed deviations from MOST even over a wide valley floor, while \cite{marti2022flux} found disagreement with MOST very close to the surface in periods of high soil moisture and high solar irradiance. 
On a steep alpine slope \cite{nadeau2013similarity} found local scaling to be inadequate due to the presence of katabatic flows and the corresponding decrease of wind speed and change of sign of the vertical wind shear above the katabatic jet maximum. Several efforts, however, have also pointed to a possibility of extending MOST scaling to complex terrain. For example, \cite{hang2021local} showed that non-dimensional wind shear below the katabatic jet can be successfully scaled if the stability parameter is corrected to include the tangent of the slope angle, while \cite{barskov2018applicability} derived a semi-empirical length scale that includes the effects of canopy and orography to extend MOST to a forested hill. 

Recently, \cite{stiperski18}, \cite{stiperski19}, \cite{stiperski21} and \cite{stiperski21prl} showed that turbulence anisotropy encodes the complexity of the boundary conditions forcing the turbulent flow, and thus can be leveraged as a missing parameter in MOST to account for the flow conditions violating MOST assumptions. \cite{stiperski18} and \cite{stiperski19} showed that anisotropy explains $80\%$ of the scatter in the scaling relations of velocity and temperature variances, as well as the scaled turbulence kinetic energy (TKE) dissipation rates observed both over flat and highly complex terrain. Based on these results, \cite{stiperski2023generalizing} developed a first generalized extension of MOST, including turbulence anisotropy as an additional non-dimensional parameter into the flux-variance relations, that proved successful over a wide range of terrain types (from flat to mountain top) and stratifications ($|\zeta| >> 1$) - conditions where basic assumptions of MOST are clearly violated and MOST is known to fail. Indeed, the importance of accounting for the anisotropy of turbulence eddies was already highlighted by \cite{katul2011mean} and \cite{salesky2013buoyancy} among others when developing the analytical model for $\phi_M$ from a modified O'KEYPS equation \citep{li2012mean}. The promising new approach of including anisotropy into scaling \citep{stiperski2023generalizing} shows the potential to cover the gap between the idealized, flat and homogeneous and narrow stability range results of the Kansas experiment and complex terrain scaling, however, its applicability to the flux-gradient relations, used in all ESMs to parametrize the surface-atmosphere exchange, has not been explored thus far.
While it is arguable that the embedded simplified assumptions of MOST remain justified over the traditional coarse resolution, characteristic in numerical models until recent years, with the continuous increase in computing power and the progressive refinement of grid resolutions this approach is not necessarily valid any longer \citep{stoll2020large}. This is especially the case in configurations with heterogeneous surface conditions, where intrinsic spatial averaging is now challenged \citep{bouzeid2020persistent}. Therefore, developing a generalized MOST scaling with extended applicability to ASL flows in both idealized and perturbed surface conditions is critical for the ABL research and modeling communities.
The purpose of this paper is therefore to extend the MOST flux-gradient scaling relations to a wide range of terrain conditions and the entire measurable stability range by including the information on turbulence anisotropy. 

The paper is organized as follows: Section \ref{sec:methods} introduces the measure of turbulence anisotropy, the datasets used and the methods employed to analyze them. Sections \ref{sec:results} and \ref{sec:Genearlization} present the state of the art scaling relations and their extension using turbulence anisotropy. In Section \ref{sec:discussion} the implications of the novel scaling relations, the limits of their validity and future applications into modelling are discussed, while conclusions are presented in Section \ref{sec:conclusions}.

\section{Data and methods}
\label{sec:methods}
\subsection{Turbulence Anisotropy in the Atmospheric Boundary Layer}\label{sec:anisotropy}
Isotropy is a mathematical symmetry defined as the invariance to rotation and reflection.  In the definition by \cite{kolmogorov1941local}, a turbulent flow is locally isotropic if the statistical properties (i.e. probability distributions) of the flow's velocity fluctuations and their products at a given point are independent of the spatial and temporal coordinates (homogeneous) and independent from any rotation and reflection of the coordinate system. Isotropy constitutes one of the pillars of turbulence theory and is often taken to be valid at small scales (inertial and dissipation ranges) where turbulence is assumed to have forgotten its generation by anisotropic boundary conditions \citep{kolmogorov1941local}. At the larger, energy containing scales, turbulence is anisotropic due to the anisotropic nature of the external forcing acting preferentially in specific directions. For example, in shear driven turbulence energy is predominantly injected into the streamwise component, while in buoyancy driven turbulence it is in the vertical component. Alternatively, wall blocking, as well as stable stratification cause preferential energy reduction in the vertical direction \citep{wyngaard2010}. 

While many studies of ABL turbulence employ the aspect ratio of velocity variances as a proxy for turbulence anisotropy \citep[e.g.][]{zilitinkevich2013hierarchy}, here we quantify it through the analysis of the Reynolds stress tensor's invariants  \citep[cf.,][]{pope2000turbulent}. The anisotropy invariants are derived from the normalized anisotropic contribution of the Reynolds stress tensor, the Reynolds anisotropy tensor
\begin{equation}\label{eq:anisReynolds}
b_{ij}=\frac{\overline{u_i' u_j'}}{\overline{u_i' u_i'}}-\frac{1}{3} \delta_{ij}.
\end{equation}

An infinite number of sets of two invariants can be derived from a purely anisotropic, therefore traceless, symmetric tensor \citep[cf.,][]{lumley1977return}. Following \cite{stiperski2023generalizing}, here we use the barycentric set of invariants $(x_b,y_b)$ as introduced by 
\cite{banerjee07}, defined as

\begin{equation}
\begin{aligned}
    x_b &=\lambda_1-\lambda_2+\frac{1}{2}(3\lambda_3+1)\\
    y_b &=\frac{\sqrt{3}}{2}(3\lambda_3+1).
\end{aligned}
\end{equation}
where $\lambda_i$ are the eigenvalues of the anisotropy tensor $b_{ij}$ \citep{pope2000turbulent}, which satisfy that 

\begin{equation}
\sum_{i=1}^{3} \lambda_i = 0 \quad \text{and} \quad \lambda_1 > \lambda_2 > \lambda_3.
\end{equation}

This barycentric mapping of the eigenvalues organizes all realizable states of turbulence into a triangular, equilateral region and gives equal weight to all of the limiting states of anisotropy and is thus preferred over the principal set of invariants as originally introduced by \cite{lumley1977return}.

Studying the eigenvalues of $b_{ij}$, the rank of $\overline{u_i'u_j'}$ and its symmetries, allows identifying three asymptotic states that correspond to the three vertices of the barycentric map \citep{pope2000turbulent}. For three component (isotropic) turbulence, $\overline{u_i'u_j'}$ has rank three (three non-zero eigenvalues), its eigenvalues are equal, while $b_{ij}$ is the null tensor and has all zero eigenvalues. This limiting case is represented by the top corner $(0.5,\sqrt{3}/2)$ of the barycentric map. For 
two component (axisymmetric) turbulence, $\overline{u_i'u_j'}$ has rank two (two non-zero eigenvalues) and is invariant through rotations and reflections around an axis, in which case the two non-zero eigenvalues of $\overline{u_i'u_j'}$ are equal and the eigenvalues of $b_{ij}$ are $\lambda_{1} = \lambda_{2} = 1/6$ and $\lambda_3 = -1/3$. This state corresponds to $(x_b,y_b) = (0,0)$, bottom-left corner. For one component turbulence $\overline{u_i'u_j'}$ has rank one (one non-zero eigenvalue), while the eigenvalues of $b_{ij}$ are $\lambda_{1} = 2/3 \text{ and } \lambda_{2,3}=-1/3$. This state is represented in the bottom-right corner of the baricentric map: $(1,0)$.

The $x_b$ coordinate thus spans from the two component axisymmetric $(x_b = 0)$ to the one component $(x_b = 1)$ limit, representing information on the componentiality of $\overline{u_i' u_j'}$, while the $y_b$ coordinate ranges from anisotropic $(y_b = 0)$ to isotropic turbulence $(y_b = \sqrt{3}/2)$. 

Following earlier work of \cite{stiperski19} and \cite{stiperski2023generalizing}, the second  coordinate of the barycentric map ($y_b$) will be used as a representative measure of turbulence anisotropy (referred hereon as the \textit{degree of anisotropy}), and will be introduced as an additional non-dimensional variable in the traditional Monin-Obukhov scaling framework of the atmospheric surface layer to generalize the scaling of the surface gradients.

\subsection{Datasets and postprocessing}

To study the flux-gradient relations over a wide range of terrain conditions, data from four field campaigns, including five multilevel meteorological towers equipped with high frequency sonic anemometers and low frequency thermocouples, was analyzed. Table~\ref{tab:datasets} contains relevant information on the datasets used.

The Cooperative Atmosphere $-$ Surface Exchange Study 1999 \citep[CASES-99;][]{poulos2002cases}, the Advection Horizontal Array Turbulence Study \citep[AHATS;][]{salesky2013buoyancy} and the NEAR tower from the Second Meteor Crater Experiment \citep[METCRAX II; ][]{lehner2016metcrax}, represent conditions close to the ideal horizontally homogeneous and flat terrain. The NEAR tower, however, had a non zero slope angle allowing deep katabatic flows to form \citep{stiperski20}. In addition, the Central and West towers from the Terrain-induced Rotor Experiment \citep[T-Rex;][]{grubivsic2008terrain} represent more complex terrain. T-Rex took place in the Owen Valley, a long straight valley with a well-defined sharp ridge on the west side. The Central tower was located at the valley floor while the West tower was placed on the western alluvial slope.

\begin{table}
\centering
\caption{Information on the datasets used.}
\vspace*{5mm}

\resizebox{\textwidth}{!}{

\begin{tabular}{llllllllll}
\toprule
    \large Dataset name & \large Experiment & \large Location & \large Terrain & 
    \large\begin{tabular}{@{}l@{}}Slope\\ angle\end{tabular}  & \large Surface & \large\begin{tabular}{@{}l@{}} 3D sonic \\ anemometer \\heights [m] \end{tabular}
    &\large \begin{tabular}{@{}l@{}} Thermocouple\\ heights [m] \end{tabular}&
    \large Time extension\\
\midrule
    AHATS & AHATS & \begin{tabular}{@{}l@{}} Kettleman City,\\California  \end{tabular} & \begin{tabular}{@{}l@{}}Flat and\\ homogeneous \end{tabular}&
    $<0.5$ & \begin{tabular}{@{}l@{}}Short grass\\ stubble\end{tabular} & \begin{tabular}{@{}l@{}}1.55, 3.3, 4.24,\\ 5.53, 7.08, 8.05\end{tabular} &
    \begin{tabular}{@{}l@{}} 1.51, 3.26, 4.23,\\ 5.47, 7.01, 8.01 \end{tabular}&
    \begin{tabular}{@{}l@{}} 2 months,\\ Jun-Aug 2008 \end{tabular}
    \\[10pt]
    CASES99 & CASES-99 & Kansas & \begin{tabular}{@{}l@{}}Flat and\\ homogeneous \end{tabular}& 
    $<0.5$ & Grassland & \begin{tabular}{@{}l@{}}5, 10, 20,\\ 30, 40, 50,\\ 55 \end{tabular}&
    \begin{tabular}{@{}l@{}}5, 10, 20,\\ 30, 40, 50,\\ 55 \end{tabular}&
    \begin{tabular}{@{}l@{}} 3 weeks, \\Oct 1999 \end{tabular}
    \\[10pt]
    METCRAX & \begin{tabular}{@{}l@{}} METCRAX II \\ NEAR \end{tabular} & \begin{tabular}{@{}l@{}}Meteor Crater,\\ Arizona\end{tabular} & Gentle slope &
    $1$ & Desert & \begin{tabular}{@{}l@{}}3, 10, 15,\\ 20, 25, 30,\\ 35, 40, 45,\\ 50 \end{tabular}& 
    \begin{tabular}{@{}l@{}}3, 10, 15,\\ 20, 25, 30,\\ 35, 40, 45,\\ 50 \end{tabular}&
    \begin{tabular}{@{}l@{}} 1 month, \\Oct 2013 \end{tabular}
    \\[10pt]
    TREX & \begin{tabular}{@{}l@{}}T-Rex \\ Central \end{tabular} & \begin{tabular}{@{}l@{}}Owens Valley,\\ California\end{tabular} & Flat valley floor &
    $<0.5$ & Desert & \begin{tabular}{@{}l@{}}5, 10, 15,\\ 20, 25, 30 \end{tabular}& 
    5,15,30 &
    \begin{tabular}{@{}l@{}} 2 months,\\ Mar-May 2006 \end{tabular}
    \\[10pt]
    TREXW & \begin{tabular}{@{}l@{}}T-Rex \\ West \end{tabular} & \begin{tabular}{@{}l@{}}Owens Valley,\\ California\end{tabular} & Gentle slope &
    $3.25$ & Desert & \begin{tabular}{@{}l@{}}5, 10, 15,\\ 20, 25, 30 \end{tabular}&
    5,15,30 &
    \begin{tabular}{@{}l@{}} 2 months,\\ Mar-May 2006 \end{tabular}
    \\[10pt]
\bottomrule
\end{tabular}
}
\label{tab:datasets}
\end{table}

The raw 20 Hz data were double rotated at each height into the streamline coordinate system ($x$ is the streamwise, $y$ the spanwise, and $z$ the surface-normal coordinate) and block averaged to 1 min for stable and 30 min for unstable stratification, with prior linear detrending. The averaging period was determined based on the multiresolution flux decomposition \citep{howell1997multiresolution}, as motivated and used by \cite{stiperski18, stiperski19, stiperski2023generalizing} and recommended by \cite{casasanta21}. 
The mean potential temperature ($\Bar\theta$) was determined from the slow response thermocouple measurements with a measurement resolution of 1 min. The stability regime was determined from the sign of the buoyancy flux $\overline{w'\theta_v'}$, and only the profiles for which all levels exhibited the same sign of the buoyancy flux were considered in the analysis. 

To evaluate the vertical gradients required in the flux-gradient relations, the profiles of mean wind speed $\frac{d\Bar{u}}{dz}$ and temperature $\frac{d\Bar{\theta}}{dz}$ were interpolated and their derivative at each level was analytically computed. 
The mean wind speed profiles were interpolated using analytical fitting, with the function $U(z) = a + b\cdot z + c\cdot\log(z)$ constraining $a = -c \cdot \log(z_0)$. Here $z_0$ is the roughness length. In the presence of a low level jet a higher-order fitting function was used $U(z) = a + b\cdot z + c \cdot z^2 + d\cdot\log(z)$, constraining $a = -d \cdot \log(z_0)$. The use of the roughness length as a fixed fitting parameter has proven to be the best approach, reducing the scatter of the flux-gradient relations in near-neutral regime. While the fitting functions used here are very simple, they do not produce the high number of artifacts for some complex shaped low level jet cases , which would occur if higher order terms were included.
To interpolate the temperature profiles, the general additive model \citep{hastie2017generalized} using fourth degree polynomial smoothing splines was applied (python class \textit{scipy.interpolate.UnivariateSpline}). This method was chosen for its flexibility and because it leads to lower fitting error. For the T-Rex campaign datasets (TREX and TREXW), where only three levels of thermocouples were available, a simple log-linear interpolation $\theta(z) = a + b*z + c*\log(z)$ was used, filtering out cases of low level jet since they produce complex shaped temperature profiles. Smoothing splines were not used to fit the wind profiles because the method produces a suspiciously high neutral limit and more scatter, due to the tendency of the splines to overestimate the local curvature of the profiles and the gradient in the first measurement level. 

Roughness length $z_0$ for each site was estimated from logarithmic fit, $U(z) = \frac{u_*}{\kappa} \log (\frac{z}{z_0})$ to the 5-minute averaged neutral $(|\zeta| < 0.02)$ profiles and the measured $u_*$.

Data were quality controlled for stationarity and well resolved heat flux. Thus stably stratified data for which $|\overline{w'\theta'}| < 0.001 K\cdot m s^{-1}$ were filtered out \citep[cf.,][]{klipp04,nadeau2013similarity}. Stationarity was assured based on the standard test by \cite{foken1996tools} at its accepted threshold of 30\%. In stable regime, the stationarity was required for the mean wind speed, buoyancy flux and momentum flux. Afterwards, consecutive one-minute-averaged profiles of fluxes and mean variables were re-averaged to 10 minutes intervals to reduce the random error. In the unstable regime, stationarity of the mean wind speed and buoyancy flux were required with a 30\% threshold; stationarity of the momentum flux was left unconstrained due to low values of the momentum flux in the convective regime.

The flux and gradient Richardson numbers
\begin{equation} 
\begin{aligned}
Ri_f &=
\frac{g}{\Bar{\theta}}\frac{\overline{w'\theta'}}{\overline{u' w'}\frac{d\Bar{u}}{dz}}, \\
    Ri &=
\frac{g}{\Bar{\theta}}\frac{\frac{d\Bar{\theta}}{dz}}{\frac{d\Bar{u}}{dz}^2},
\end{aligned}
\end{equation} 
were required to be below 0.25 \citep[cf.,][]{grachev2013critical,babic2016evaluation}. In case of a low level jet, only the wind speed gradient from measurements below the jet maximum height were considered. Additionally, cases of counter-gradient fluxes where $\overline{u' w'}d\Bar{u}/dz > 0 $ or $\overline{\theta' w'}d\Bar{\theta}/dz > 0 $, were filtered out.

In the stable regime, applying stationarity criteria reduces the number of 1-min periods to 40\% (averaging through the five datasets), with TREXW tower being the most affected dataset (20\% surviving). After re-averaging the stationary data to 10 minutes blocks, the Richardson number criteria leaves 70\% of data points available and the number of non-countergradient cases averages to 50\% for temperature and 60\% for wind profiles surviving the previous criteria. On average one third of the stationary re-averaged data points is present in the quality controlled data. In the unstable regime, where the averaging period is 30 minutes, the impact of the quality criteria is much lower, with 90\% of stationary points and 80-90\% of down-gradient cases.

\begin{table}
\caption{ Number of data points and relative percentage population of the ensemble dataset in the two stability regimes.}
\vspace*{5mm}
\centering
\begin{tabular}{ccccccc}
\toprule
    regime & data points & AHATS & CASES99 & METCRAX & TREX & TREXW
    \\
\midrule
    Stable & 36765 & 28.4 & 9.0 & 18.8 & 28.9 & 14.9 \\
    Unstable & 24008 & 25.6 & 7.2 & 18.6 & 23.3 & 25.2\\
\bottomrule
\end{tabular}
\label{tab:ensemble}
\end{table}

\subsection{Formulating new scaling relations}
\label{sec:meth_scal} 

To create new parametrizations including $y_b$ in the flux-gradient relations, we used local scaling, as is common in other scaling studies \citep[cf.,][]{stiperski2023generalizing}. The high quality data from each dataset was first merged into an ensemble further used for the analysis. The total number of data points and the relative percentages of points in each individual datasets are shown in Table~\ref{tab:ensemble}. Fitting the model directly to this ensemble allows the analysis to cover a variety of terrain types and meteorological conditions, with the best model being a reasonable estimate of the average dependence of the scaling relations on stability and anisotropy. This method, however, leads to an uneven representation of each dataset in the ensemble, due to the different amount of data points per site, and does not allow assessing inter-dataset differences. Thus, the robustness of this method will be addressed below.

The new parametrizations were obtained in the following way. For each target variable ($\phi_M, \phi_H$) and separately for the unstable and stable regime, a basis curve was chosen among the ones available in literature (cf., Tables \ref{tab:unstablescaling} and \ref{tab:stablescaling}) based on its ability to reproduce the evident features of the data in an optimal way. The bi-variate robust fit was then used to fit the chosen curve to the ensemble data, and determine the coefficients ($\alpha_i$) of the model. Since the dependence of the model coefficients on anisotropy has to be prescribed, we first determined whether the model should have a linear or a non-linear dependence on $y_b$ by dividing the data into bins of $y_b$ and fitting the curve to each bin, and then examining the resulting $\alpha_i(y_b)$ curves, as in \cite{stiperski2023generalizing}. Once the best functional form of $\phi(\zeta,y_b)$ was chosen for each case, the coefficients of the obtained model were tuned by minimization of the Huber loss function \citep[][using the python function \textit{scipy.optimize.least\_squares} with loss = 'huber']{huber1992robust}, a more robust alternative to the ordinary least squares method.

The performance of the model was compared to that of the HÖ96 parametrizations, using the skill score defined as:
\begin{equation}
\label{eq:MAD}
    SS = 1 - \frac{MAD_{new}}{MAD_\textit{HÖ96}}
\end{equation}
where the MAD is the median of the absolute value of the logarithmic residuals $\{|\log(\phi_{data})-\log(\phi_{model}(\zeta,y_b))|\}$, $MAD_\textit{HÖ96}$ is the MAD using the HÖ96 parametrizations and $MAD_{new}$ is the MAD of the new model including the dependence on $y_b$.

To assess if the differences in representation between datasets could influence the result, each model (separately for stable and unstable, windspeed and temperature) was tested for compatibility against the average model between the five datasets. To do so, the ensemble data was bootstrapped a thousand times with replacement into groups of half the number of points of the original ensemble dataset and the model $\phi(\zeta,y_b)$ refit to each sample; for every parameter $\alpha_{i,ens}$ of the model, the standard deviation $\sigma_{i,ens}$ was calculated. Then, the same model was trained on each of the five datasets separately and the average $\alpha_{i,avg}$ between the five estimates was calculated for each parameter. If the difference in the weight of each dataset does not influence the results, the average model between the five datasets and the ensemble data model would be the same. To test this hypothesis, Student's t-test was used on every parameter, comparing the ensemble data estimate and the average-of-five-datasets estimate, using a 5\% significance level: $t_i = \dfrac{|\alpha_{i,ens}-\alpha_{i,avg}|}{\sigma_{i,ens}} < 1.96$. Results of this procedure are discussed in Section~\ref{sec:discussion}.

Finally, to assess inter-dataset differences, a-posteriori leave-one-out crossvalidation was applied to each model. This procedure consists in training the model on a subsample of data containing all datasets minus one and validated on the one left out of training calculating the skill score. This allows to assess  differences between the datasets and reduce the bias in the assessment of the overall performance that is caused by validating a model on the same data on which it is trained.

For visualization purposes, when plotting the scaling relations, data are binned into logarithmically distributed bins of $\zeta$, defined through the local Obukhov length.
This choice does not affect the analysis but gives a better visualization of the data, since $\zeta$ is exponentially distributed.

\section{Flux-gradient scaling relations and the role of turbulence anisotropy}
\label{sec:results}
\begin{figure}
\centering
\includegraphics[width=0.8\textwidth]{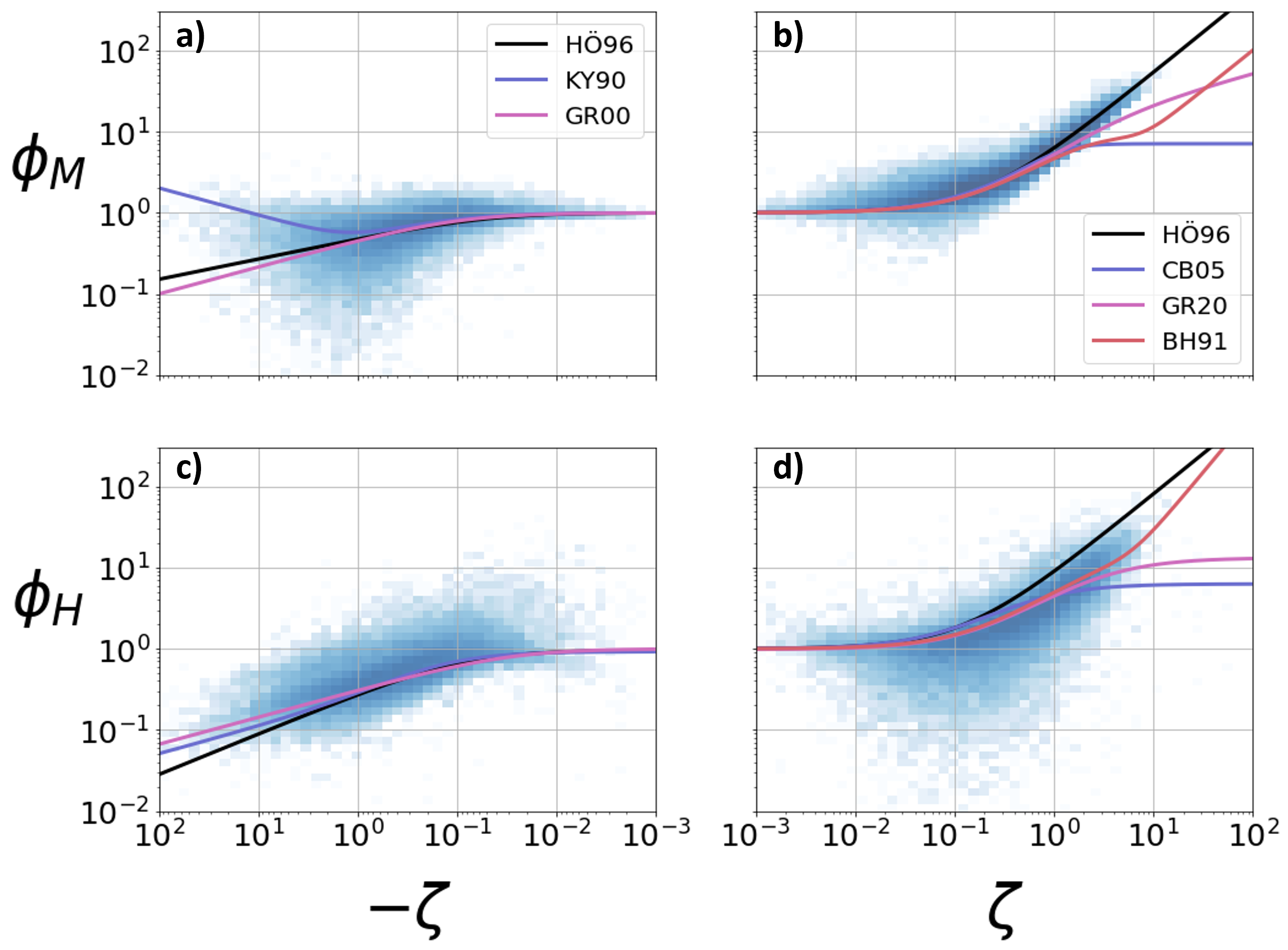}
\caption{Two-dimensional histogram of flux-gradient relations for the ensemble dataset for the mean wind speed (a,b) and temperature (c,d) profile under unstable (a,c) and stable (b,d) stratification. Literature curves from Tables~\ref{tab:stablescaling}~and~\ref{tab:unstablescaling} are shown in color.}
\label{fig:scaling}
\end{figure}

We next explore the validity of the universal scaling relations for the windspeed and temperature profiles. 
As illustrated in Figure \ref{fig:scaling}, the literature $\phi_M$ curves (Table \ref{tab:unstablescaling}) in the unstable regime for $\zeta > -0.5$ show good agreement amongst themselves and are supported by the data. For $\zeta < -0.5$, however, the data scatter increases and it is unclear which power law the data are following. The presence of substantial scatter in this regime indicates the existence of a missing explanatory variable as argued by \cite{salesky2012random}. The disagreement also partially stems from the fact that the classical studies leading to the accepted scaling relations only considered data up to $\zeta \approx - 2$, while the range of data used in this study extends down to $\zeta = -100$.
Under stable stratification, the wind speed gradient $\phi_M$ is well represented by a linear function of $\zeta$, as in the scaling relations of \cite{hogstrom96}. \cite{grachev2013critical} and \cite{babic2016evaluation} showed that this is a consequence of filtering out data with a supercritical Richardson numbers, which would otherwise show a leveling off at high stabilities.

The scaled potential temperature gradient $\phi_H$ in the unstable regime is overall higher than predicted by classical scaling curves (Table \ref{tab:unstablescaling}). Due to large scatter it is also unclear which power law the data are following. On the other hand, the stable regime $\phi_H$ could also be well represented by a linear model, as argued for $\phi_M$, however, it is clear that the HÖ96 curve is not representative of the available data. Furthermore a great amount of scatter is present around $\zeta = 0.1$.

\begin{figure}
\centering
\includegraphics[width=0.8\textwidth]{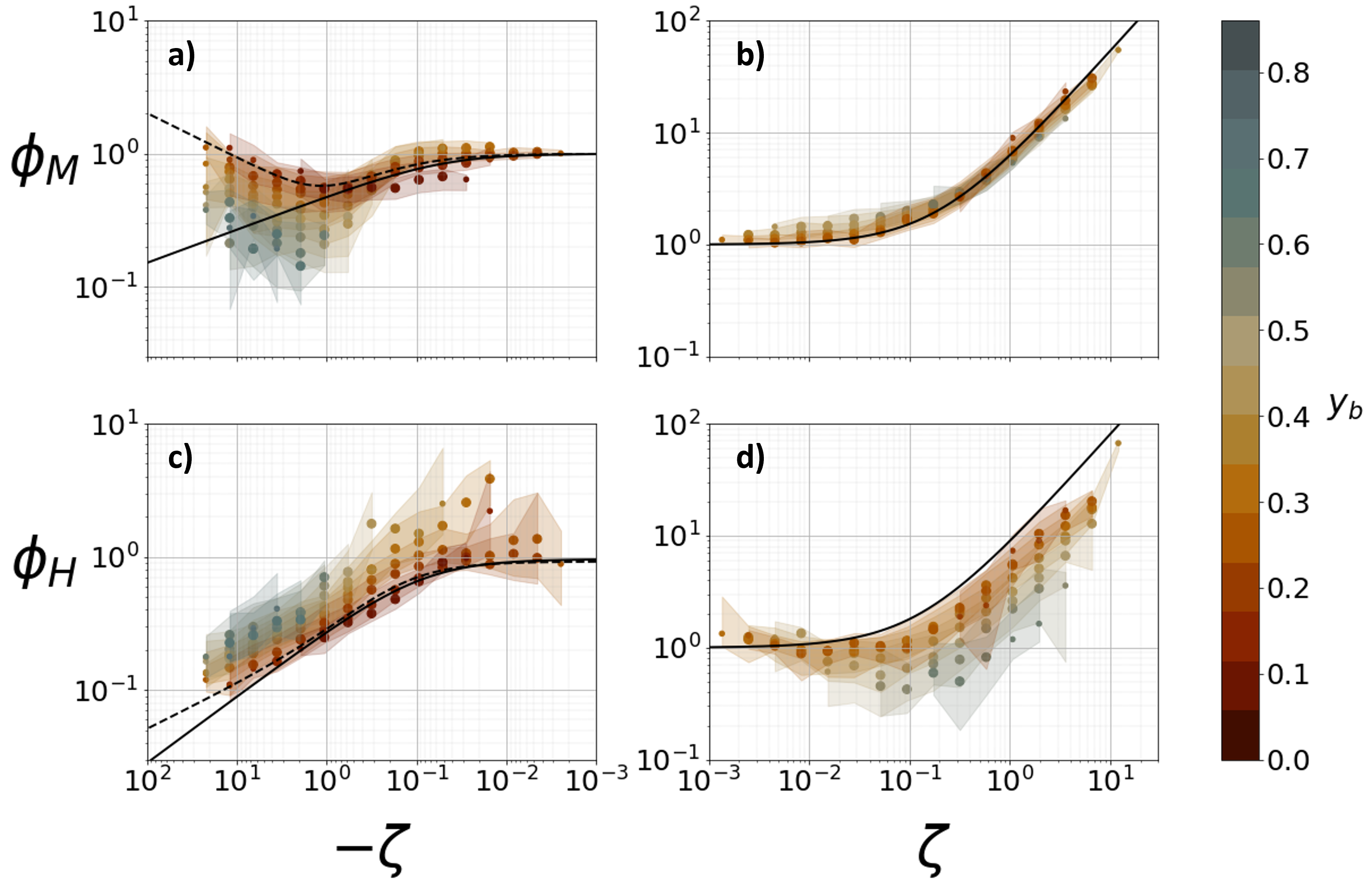}
\caption{ As in Figure \ref{fig:scaling} but for data binned according to the degree of anisotropy $y_b$ (color bar). 
The black lines represent the HÖ96 parametrizations, while the dashed lines in the unstable regime represent the KY90.}
\label{fig:ybbinned}
\end{figure}

In order to further investigate the observed scatter and to understand the behavior of the dimensionless gradients, we next bin the data according to the degree of turbulence anisotropy, $y_b$ (Figure~\ref{fig:ybbinned}). 
Once the data are clustered based on the degree of turbulence anisotropy, different scaling behaviors emerge, suggestive of the relevance of turbulence anisotropy as an additional independent parameter, missing in traditional MOST scaling. In the unstable regime this parameter's variability accounts for a significant portion of the observed scatter. In fact, the data show that for anisotropic turbulence the normalized velocity gradient follows the scaling behavior of \cite{kader90} closely, with a $-\zeta^{1/3}$ power law, while for more isotropic turbulence, the scaling curves follow the same power law but with progressively lower multiplication coefficients. 
The debate between the existence of a $1/3$ versus $-1/3$ power law in the free convective limit, where $\zeta< -1$, has been ongoing since the 1980s \citep{hogstrom88}. While the $-1/3$ power law can be inferred from analytical arguments \citep[namely the O'KEYPS equation,][]{li2021keyps}, no analytical model has been able to predict the $+1/3$ behaviour, which can nevertheless be derived by directional dimensional analysis \citep{kader90} and scaling arguments. In fact, in the free convective limit where $u_*$ is no longer a relevant parameter, having $\phi_M\sim -\zeta^{\frac{1}{3}}$ will cancel out the dependence on friction velocity on both sides of the equation. The same result is achieved by considering the free convection velocity $w_{fc} = \left[\frac{g}{\theta}\overline{w'\theta'}z\right]^{1/3}$ as the relevant velocity scale. As argued by \cite{li2021keyps}, however, this asymptotic behaviour $\phi_M \sim -\zeta^{1/3}$ is in contradiction with the O'KEYPS equation, which is based on the assumption of a constant asymptotic limit of the Prandtl turbulent number, $Pr_t$. Interestingly, results in Figure \ref{fig:ybbinned} illustrate that in the highly convective range, $\phi_M$ effectively increases, and thus the turbulent Prandtl will go to zero as showed in Section~\ref{sec:ks}, highlighting the ineffectiveness of the O'KEYPS model when $\zeta < -1$.
In contrast, $\phi_M$ in the stable regime shows little dependence on anisotropy. 

In parallel, turbulence anisotropy explains substantial scatter of $\phi_H$ under both stratifications. In the unstable regime similarly to $\phi_M$, $\phi_H$ shows a clear dependence on $y_b$, with anisotropic data following the scaling behaviour of \cite{kader90}, while isotropic data are parallel but with higher values. In the the stable regime, anisotropy accounts for the majority of the scatter in $\phi_H$, but the scaled data highlight the need to correct the slope of the HÖ96 parametrization. 

\section{Generalized parametrization of the flux-gradient scaling relations}\label{sec:Genearlization}

\subsection{Unstable regime}

As shown in Sect. \ref{sec:results}, the $\phi_M$ in the unstable regime binned according to anisotropy $y_b$ (Figure~\ref{fig:scaling}a) clearly support the model of \cite{kader90}, which we use to develop the new scaling relations including anisotropy.
The KY90 model exists in two functional forms as introduced in \cite{kader1989effect} and \cite{brutsaert1992stability}. We follow the \cite{brutsaert1992stability} formulation due to its simpler form that allows easier interpolation to the ensemble data (cf. Table \ref{tab:unstablescaling}):

\begin{equation}
\label{eq:unstablemodelM}
    \phi_M = \frac{ a(y_b) + b |\zeta|^{n(y_b)}}{a(y_b) + |\zeta|^{n(y_b)}} - c(y_b) \cdot \zeta^{\frac{1}{3}}.
\end{equation}
Here $a,n$ and $c$ are the model coefficients ($\alpha_i$) and are linear functions of $y_b$. The procedure for determining the best model and tuning it is described in Section~\ref{sec:methods}.

Fitting the model results in the following relations:
\begin{align*}
    a &= \begin{cases}
        0.24 - 0.38 y_b & y_b<0.6\\
        0.012 & y_b >0.6,\\
    \end{cases}\\
    b &= 0.061,\\
    c &= 0.45 - 0.53 y_b,\\
    n &= -0.12 + 6.4 y_b,
\end{align*}
where the parameter $a$ is constrained to be positive.
The resulting curves for each anisotropy bin are shown in Figure~\ref{fig:Model}a.

Similar results are obtained for $\phi_H$ in the unstable regime. When $\phi_H$ is binned according to anisotropy (Figure~\ref{fig:scaling}c) it clearly shows that the KY90 formulation should be employed. Of the two functional forms following the \cite{kader90} model, the one from \cite{kader1989effect} is chosen here, since the function provided by \cite{brutsaert1992stability} has a constant asymptotic limit that does not agree with the power law predicted in the convective sub-layer ($\zeta^{-1/3}$), as supported by the data. 
The chosen model then takes the form:
\begin{equation}
\label{eq:unstablemodelH}
    \phi_H = a(y_b) \left[ \frac{3 - 2.5\zeta}{1 - 10 \zeta + 50 \zeta^2 }  \right]^{\frac{1}{3}}
\end{equation}
with $a = a_0 + a_1 y_b$. It is estimated that $a_0 = 0.48 \text{ and } a_1 = 1.8$. The model is shown in Figure~\ref{fig:Model}c.

\subsection{Stable regime}

\begin{figure}
\centering
\includegraphics[width=0.8\textwidth]{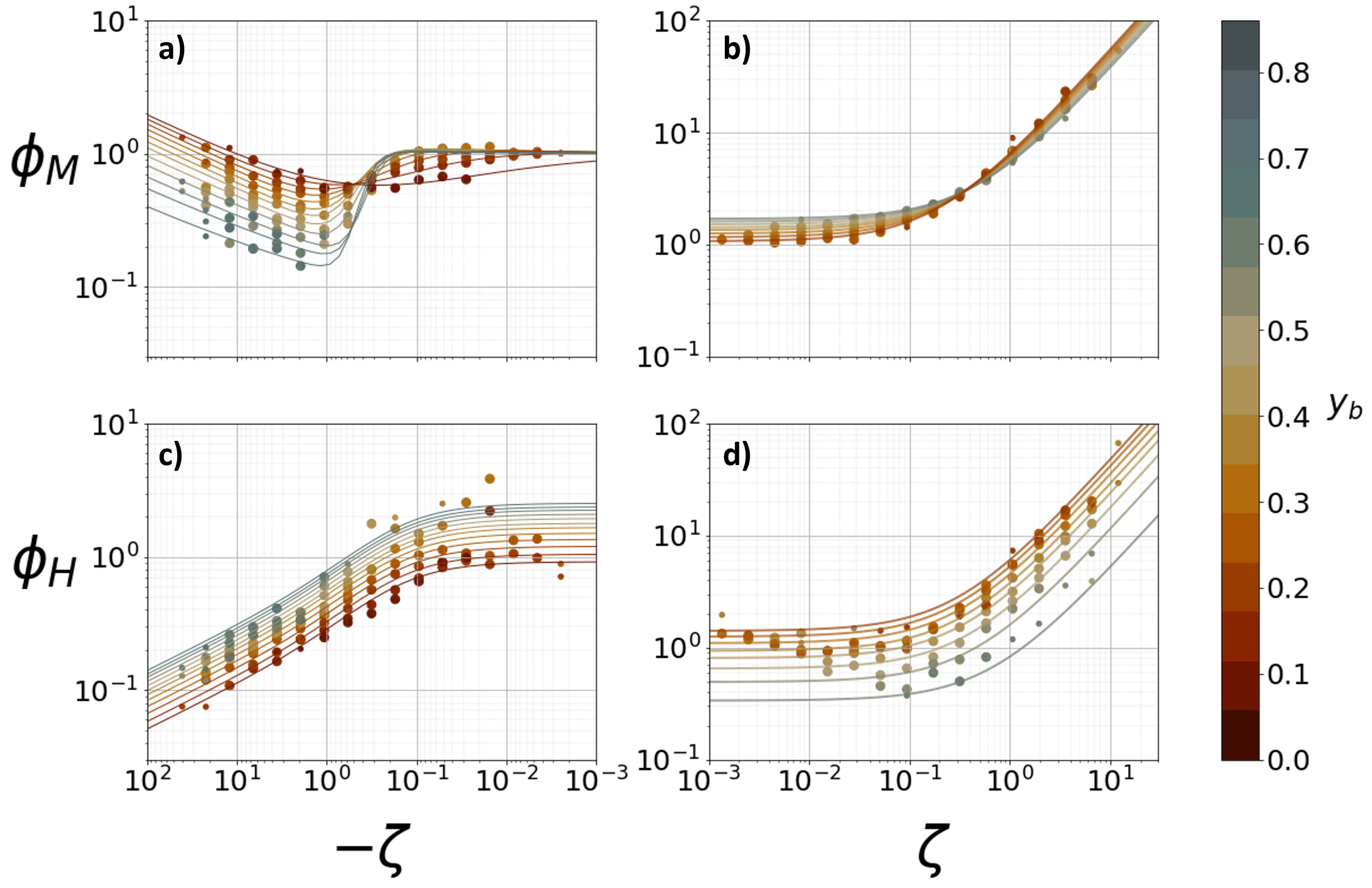}
\caption{The extended flux-gradient universal scaling relations that include the effect of turbulence anisotropy through the degree of anisotropy parameter $y_b$. Data are binned according to the $y_b$ value (red is anisotropic and dark gray is isotropic turbulence) and $\zeta$, then binned medians are plotted. The corresponding scaling curve for each $y_b$ bin is plotted on top of data. The organization of the panels is the same as in Figures~\ref{fig:scaling} and \ref{fig:ybbinned}.}
\label{fig:Model}
\end{figure}

Given the historically accepted linear nature of the flux-gradient relations in the stable regime (when $Ri$ and $Ri_f$ are below 0.25), also supported by the ensemble data, as well as the role of turbulence anisotropy illustrated in the previous section, we use here a multilinear regression with $\zeta$ and $y_b$ as predictors:
\begin{equation}
\label{eq:stablemodel}
    \phi = a(y_b) + b(y_b) \zeta,
\end{equation}
where the parameters $a$ and $b$ are linear functions of anisotropy. The parameters are fit to data for $\phi_M$ (Figure~\ref{fig:Model}b):
\begin{equation}
\begin{aligned}
        a &= 0.76 + 1.5 y_b,\\
        b &= 6.3 - 4.3 y_b,\\
\end{aligned}
\end{equation}
and for $\phi_H$ (Figure~\ref{fig:Model}d):
\begin{equation}
\begin{aligned}
        a &= \begin{cases}
                1.9 - 2.6 y_b & y_b<0.6\\
                0.34 & y_b>0.6\\
            \end{cases},\\
        b &= 6.7 - 10 y_b.\\
\end{aligned}
\end{equation}
where $a$ is again constrained to be positive.

\subsection{Implications of the new scaling relations}

The new flux-gradient relations for windspeed and temperature highlight a number of interesting features. All the parametrizations show that more anisotropic turbulence is closer to the existing scaling relations, while the quasi-isotropic turbulence shows large deviations. In particular, the parametrization for $\phi_M$ in the unstable regime supports the existence of three sub-layers (dynamic, dynamic-convective and convective) introduced by KY90, and shows that the scaling curves follow the general scaling prescribed by \cite{kader90}, however, the ranges of $\zeta$ for each sub-layer depend on turbulence anisotropy. Thus, the more anisotropic turbulence is closest to the original KY90 curve, and shows the onset of the dynamic-convective sub-layer at smaller $|\zeta|$ than the more isotropic turbulence. In fact, the scaling relations suggest a regime transition at around $-\zeta \sim 0.4$ so that in dynamic sub-layer $\phi_M$ is larger for less anisotropic turbulence, whereas in dynamic-convective and convective sub-layers the opposite is the case. Interestingly, the same is true also in the stable regime, where $\phi_M$ shows a dependence on anisotropy that changes at around $\zeta \sim 0.3$. Thus  $\phi_M$ is proportional to $y_b$ in dynamic regime where buoyancy effects are of lesser importance, and decreases with increasing $y_b$ in conditions when buoyancy cannot be neglected. 

For the temperature gradient the results are different, and there is no clear indication of regime transitions. However, the results show the opposite impact of anisotropy in unstable and stable stratification. In the unstable regime $\phi_H$ is proportional to $y_b$ and thus a higher $\phi_H$ is observed for isotropic turbulence, whereas in stable stratification more isotropic turbulence occurs with smaller $\phi_H$.
In the stable regime the slope of the stability correction is strongly modulated by $y_b$, with the limit of a zero slope for very isotropic turbulence ($y_b > 0.6$). Finally, it is important to note that the neutral limit of the temperature parameterizations is not continuous between stable and unstable regime, unless only anisotropic turbulence is considered.

\section{Discussion and implications for Earth System Models}
\label{sec:discussion}

\subsection{Model performance and crossvalidation}
\label{sec:crossval}
To assess the improvement in the performance of the newly developed flux-gradient scaling relations for windspeed and temperature (Equations \ref{eq:unstablemodelM},\ref{eq:unstablemodelH},\ref{eq:stablemodel}), we calculate the skill score (see Sec.~\ref{sec:meth_scal}) in comparison to the classic Businger-Dyer relations, as refitted by HÖ96. 
 
The skill score is assessed separately for each non-dimensional gradient and for the stable and unstable regimes first, and then by separating data into the very unstable ($\zeta < -0.1$), near-neutral unstable ($-0.1 <\zeta < 0$), near-neutral stable($0<\zeta<0.1$) and very stable ($\zeta>0.1$) regimes (Table~\ref{tab:performance}). 
The results show a clear improvement brought about by extending flux-gradient scaling relations to include anisotropy. Particularly large improvement is observed in unstable stratification, where the new model consistently outperforms HÖ96 both for wind speed and temperature. In this regime the biggest improvement is seen for temperature under very unstable conditions. In the stable regime the best improvement for wind speed is in the neutral regime, while the opposite happens for temperature. 
For the stable regime $\phi_H$, comparing the model to the HÖ96 parametrizations leads to a major artificial increase in performance, since the slope of HÖ96 is not supported by the data (Figure~\ref{fig:ybbinned}). Still, if the performance reference is taken as the linear model $\phi_H = a + b\zeta$ fitted to the entire ensemble dataset, the resulting skill score is 0.08, still supporting the improvement created by our model against a Businger-Dyer model.

Since each dataset had different number of data points when merged into the ensemble, and so different weights were given to each dataset, we tested every ensemble model (model for wind speed and temperature, for stable and unstable stratification) against the average model obtained for each individual dataset using Student's t-test on every model coefficient, as described in Section~\ref{sec:meth_scal}. The test was passed for every parameter for $\phi_M$ in both regimes and for $\phi_H$ in stable regime. In the unstable regime, the model for temperature lead to incompatible values, meaning that consistent dataset differences are present that do not average out in the ensemble data. This might indicate that the temperature model in the unstable regime is the least robust, or the most site dependent, of the four, or could be caused by underestimation of the parameters' variance.

\begin{figure}
\centering
\includegraphics[width=0.9\textwidth]{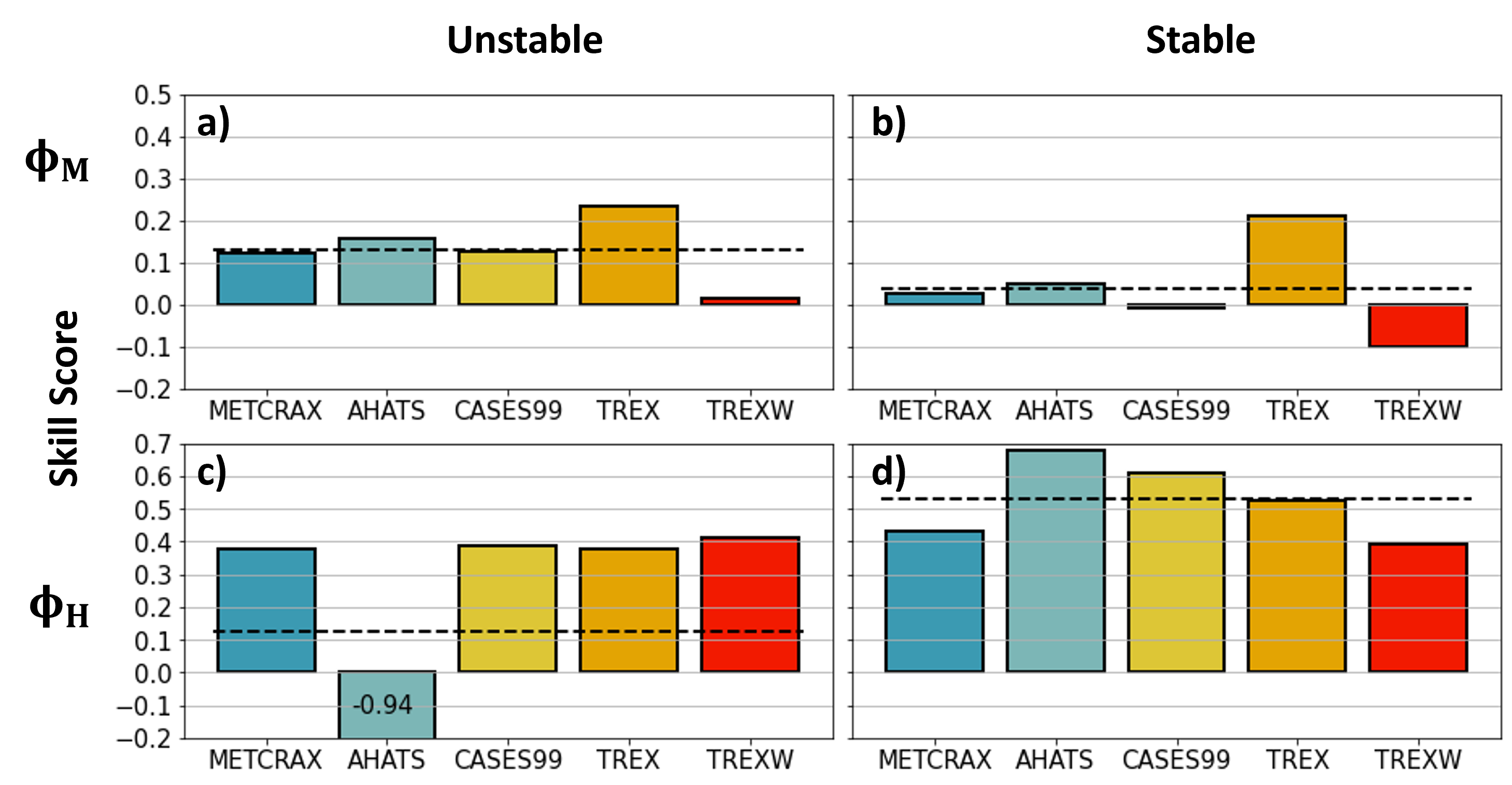}
\caption{ Individual skill score of the leave-one-out crossvalidation for $\phi_M$ (a,b) and $\phi_H$ (c,d) for unstable (a,c) and stable (b,d) stratification. Skill score is calculated as in Equation~\ref{eq:MAD}. The horizontal black line shows the average cross-validation performance.}
\label{fig:crossval}
\end{figure}

Since the new scaling relations were trained and tested on the same data, we further examine the improvement brought about by including anisotropy on each individual dataset by using a posteriori leave-one-out cross-validation (see Sec.~\ref{sec:meth_scal}). The cross-validation skill scores (Figure~\ref{fig:crossval}) show an overwhelming and consistent improvement using the new scaling relations for all stratification regimes and variables over all the datasets, as already observed for the ensemble data. Interestingly, the relatively complex terrain data at T-Rex Central tower consistently shows the largest improvement in $\phi_M$ over all other datasets. At the same time, as expected for locations with katabatic flows where $\zeta$ needs to be augumented with the information on slope angle \citep[cf.,][]{hang2021local}, the T-Rex West tower, shows a marginal degradation of performance under stable stratification.    

The performance of the new $\phi_H$ relations in unstable stratification is particularly striking (40\%), as the improvements for individual datasets are significantly larger than the overall improvement of the model (25.5\%). Generally, classic MOST $\phi_H$ relations are consistently found to underestimate the surface sensible heat flux and recently machine learning methods have been employed to reduce the error \citep{McCandless2022}. This result can clearly be connected to not separating the different anisotropy regimes when performing the analysis. 

One dataset, however, stands out in terms of reduced performance when anisotropy is taken into account, and that is AHATS, whose cross-validation skill score for  $\phi_H$ in the unstable regime is surprisingly low (Figure~\ref{fig:crossval}c). This result suggests that the HÖ96 relations outperform the new scaling relations. The potential causes of this discrepancy compared to the other datasets could stem from the fact that the AHATS tower is significantly lower than other towers (8 m compared to 30-50 m for other towers) and thus also has less isotropic data \citep{stiperski21} in unstable regime ($y_b = [0.1- 0.3]$). This inconsistency can further explain the results of the previous compatibility test. Limited improvement or slight deterioration of the results is also found for individual towers under stable stratification, where \cite{stiperski19} already showed the importance of processes other than anisotropy. 

Apart from these cases, crossvalidation shows an overall consistent improvement over classic formulations, even when these formulations were refit, which reinforces the use of the ensemble trained parametrizations of Section~\ref{sec:results} as a compromise between different sites characteristics and flow conditions.

\begin{table}
\caption{Skill score of the new scaling relation for $\phi_M$ and $\phi_H$ compared to HÖ96 model (Eq.~\ref{eq:MAD}). The skill score is calculated on the whole range of $\zeta$ first and then again separating each regime into near neutral and very (un)stable ($|\zeta|<>0.1$)}
\vspace*{5mm}
\centering
\begin{tabular}{ccc}
\toprule
     & \large Unstable & S\large table \\
     & \begin{tabular}{@{}ll@{}}$\zeta<-0.1$ & $\zeta>-0.1$ \end{tabular} &
      \begin{tabular}{@{}ll@{}}$\zeta<0.1$ & $\zeta>0.1$ \end{tabular} \\[5pt]
\midrule
    \large $\phi_M$ & 0.17 & 0.08\\
    & \begin{tabular}{@{}ll@{}} 0.15 & 0.19 \end{tabular}&\begin{tabular}{@{}ll@{}} 0.14 & 0.06 \end{tabular} \\[10pt]
    \large $\phi_H$ & 0.25 & 0.58\\
    &\begin{tabular}{@{}ll@{}} 0.27 & 0.10 \end{tabular} &\begin{tabular}{@{}ll@{}} 0.26 & 0.63\end{tabular} \\
\bottomrule
\end{tabular}
\label{tab:performance}
\end{table}

\subsection{Effect of xb in stable regime}

In developing a model for the flux-gradient relations that includes turbulence anisotropy as a scaling parameter, only the degree of anisotropy through the $y_b$ coordinate of the barycentric map (cf. Section~\ref{sec:anisotropy}) was considered, leaving out information on the componentiality of the stress tensor ($x_b$). One could, legitimately, wonder what would be the effect of including the $x_b$ coordinate into the newly developed scaling relations. The effect of $x_b$, however, can be relevant only in conditions where $x_b$ and $y_b$ are uncorrelated, which is only true for the stable regime. In fact, calculating the Pearson correlation coefficient $\rho$ between $x_b$ and $y_b$ separately for the unstable and stable regime results in $\rho = 0.4$ (unstable) and $\rho = -0.03$ (stable).

To include $x_b$ in the stable regime scaling relations, Equation~\ref{eq:stablemodel} was modified as:

\begin{equation}
\label{eq:stablexb}
    \phi = a(x_b,y_b) + b(x_b,y_b) \zeta, 
\end{equation}
where 
\begin{equation}
\begin{aligned}
        a &= a_0 + a_1 y_b + a_2 x_b,\\
        b &= b_0 + b_1 y_b + b_2 x_b,
\end{aligned}
\end{equation}
and fitted to the data for both $\phi_M$ and $\phi_H$.

Fitting the model to the ensemble data shows no improvement in the performance. In fact, the skill-score decreases by 0.002 (0.2\%) for the stable windspeed gradient model and increases by 0.005 (0.5\%) for the stable temperature gradient model. 

This analysis concludes that the effect of $x_b$ in the scaling relations for $\phi_M$ and $\phi_H$ is negligible.

\subsection{Scaling components analysis}
\label{sec:components}
\begin{figure}
\centering
\includegraphics[width=0.9\textwidth]{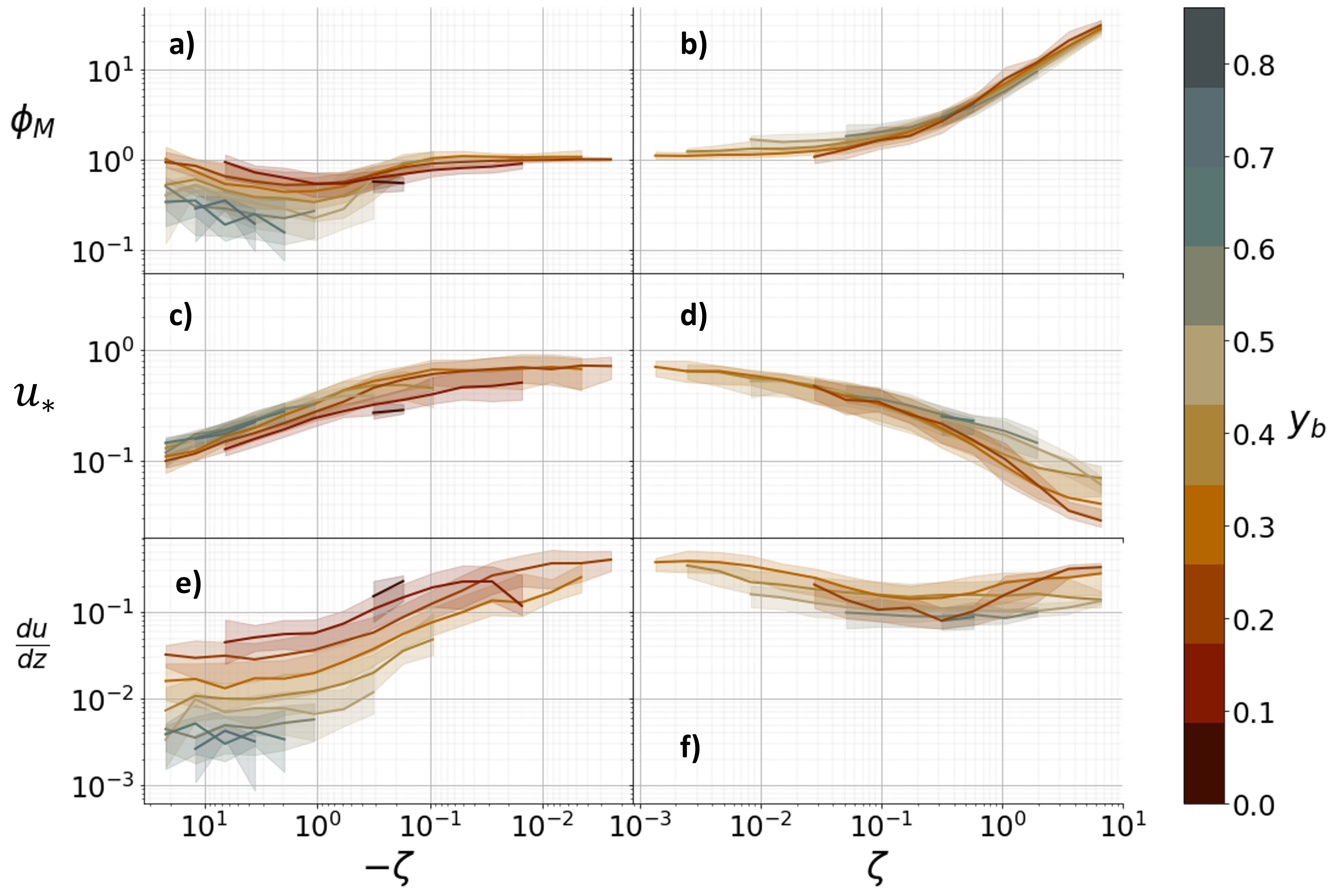}
\caption{ The components of the scaling relation for $\phi_M$ are shown with their dependence on $\zeta$ and $y_b$ for both the stable (right column) and unstable (left column) regime.}
\label{fig:components_wind}
\end{figure}

\begin{figure}
\centering
\includegraphics[width=0.9\textwidth]{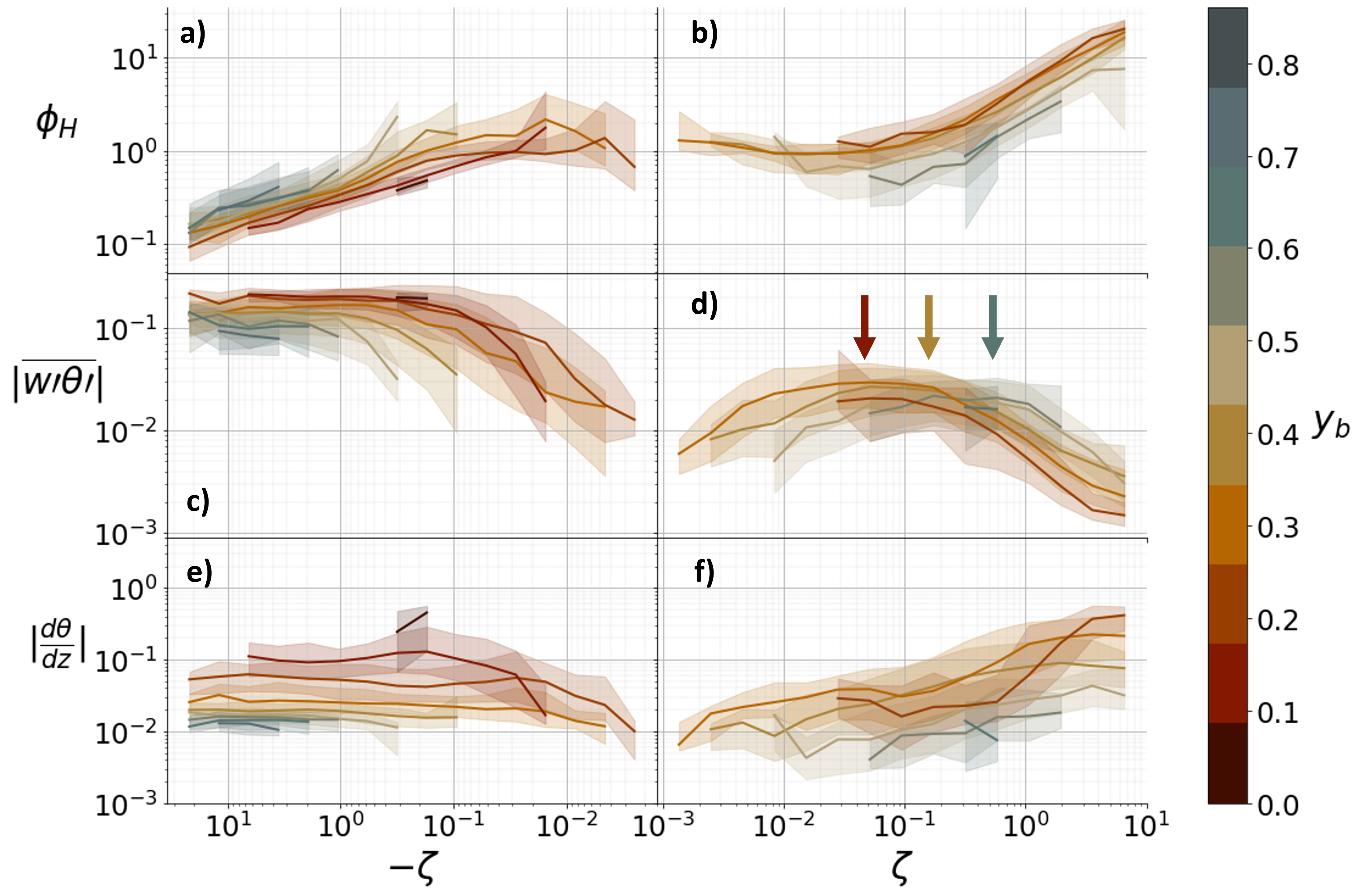}
\caption{The components of the scaling relation for $\phi_H$ are shown with their dependence on $\zeta$ and $y_b$ for both the stable (right column) and unstable (left column) regime.}
\label{fig:components_temp}
\end{figure}

While the new scaling relations show a clear stratification according to anisotropy, thus hinting at the different ABL structure and dynamics for the different anisotropy types, it still remains unclear if the fluxes or gradients are affected more by anisotropy and in what way. We therefore find it instructive to investigate the dependence of each individual dimensional component that forms each scaling relation on anisotropy. While this approach is clearly limited, since the variables are dimensional and can acquire a range of values, it can still offer insight into the ABL structure. Figure~\ref{fig:components_wind} thus shows the dependence of $\phi_M$, $u_*$ and $\frac{du}{dz}$ on $\zeta$ and $y_b$, while Figure~\ref{fig:components_temp} shows the dependence of $\phi_H$, $\overline{w\theta}$ and $\frac{d\theta}{dz}$ on $\zeta$ and $y_b$.

In the convective region of the unstable regime (Figure~\ref{fig:components_wind}a,c,e), when $\zeta \to -\infty$, $u_*$ decreases to values around 0.1, while the temperature gradient and the buoyancy flux are asymptotically constant (Figure~\ref{fig:components_temp}a,c,e), meaning that close to the surface where our data reside the increasing dynamic instability is driven by a decreasing dynamic forcing and not by an increasing buoyancy forcing. The rise of $\phi_M$ predicted by \cite{kader90} for $\zeta <-1$ and observed in our data are shown to stem from the windspeed gradient being almost constant, while the $u_*$ is monotonically decreasing. In this range, isotropic turbulence (high $y_b$, dark gray) exhibits larger momentum flux and smaller kinematic buoyancy flux, as well as smaller windspeed and temperature gradients than its anisotropic counterpart, which explains the stratification with $y_b$ observed in the convective regime (cf., Figure~\ref{fig:ybbinned}). Still, the drop in the gradients rather than the change in the fluxes in isotropic versus anisotropic turbulence appears to be the driving mechanisms in the observed dependence of scaling relations on anisotropy. 

In near-neutral unstable conditions, where $\zeta \to 0^-$, the buoyancy flux tends to zero faster then the temperature gradient in a well defined way for any degree of anisotropy, while $u_*$ approaches one, thus leading to a neutral limit: \[\lim_{\zeta \to 0^-} \phi_H(\zeta) > 1.\] A similar result was highlighted by \cite{tampieri2009investigation}, \cite{sfyri18} and \cite{stiperski2023generalizing} for the scaling of the temperature variance. This result implies a turbulent Prandtl number, $Pr_t$, greater than unity in the near-neutral unstable limit, that is an increasing function of $y_b$ (see Sec.~\ref{sec:ks}). 
In the unstable regime $\frac{d\theta}{dz}$ is constant with $\zeta$ when decomposed into anisotropy bins, while its neutral limit decrease is uncertain and not consistent. This uncertainty (also encountered in the stable neutral limit) arises from the unclear nature of the neutral limit in the surface layer and the difficulties of fitting almost constant temperature profiles. In conditions of very small $|\zeta|$ the typical profile of temperature is overall constant with height, but oscillations at the scale of ten meters without a change in the buoyancy flux sign are present, which increase the difficulty in the determination of $\lim_{\zeta \to 0}\frac{d\theta}{dz}$. This explains the difficulty in determining the unstable neutral limit for $\phi_H$ and $Pr_t$.

In the near-neutral range of the stable regime (Figure~\ref{fig:components_wind}b,d,f), where $0<\zeta<0.1$, $u_*$ and $\frac{du}{dz}$ decrease almost in parallel, while in the stable range, where $\zeta>0.1$, the wind speed gradient increases with increasing stability while $u_*$ decreases faster than in the neutral range. This behavior leads to the linear dependence of $\phi_M$ on $\zeta$. This result implies that in the very stable boundary layer wind shear is still present but turbulence suppression by buoyancy limits the surface stress $u_*$. In particular, $u_*$ exhibits a change of behaviour when $\zeta >= 0.1$ with isotropic turbulence having a larger $u_*$ and tending to zero slower than anisotropic turbulence. In this intermediate stability regime the kinematic buoyancy flux has a peak as showed by \cite{grachev05}, which happens at lower stability for anisotropic turbulence ($\zeta \sim 0.05$) than for isotropic turbulence ($\zeta \sim 0.5$), as shown by the arrows in Figure~\ref{fig:components_temp}d. The temperature gradient, however, increases with stability in the stable boundary layer, as expected. Still, isotropic turbulence exhibits lower temperature gradients than the anisotropic one, driving the behaviour of $\phi_H$.

\subsection{Eddy diffusivities and the turbulent prandtl number}
\label{sec:ks}
Eddy diffusivities for momentum, $K_m$, and heat, $K_h$, describe the rate at which turbulent eddies transport these two quantities across a fluid medium. They are a product of K-Theory \citep{stull1988introduction}, a first order turbulence closure model, formulated as:

\begin{equation}
\label{eq:K_theory}
\begin{aligned}
    \overline{u'w'} &= -K_m \frac{d\Bar{u}}{dz},\\
    \overline{w'\theta'} &= -K_h \frac{d\Bar{\theta}}{dz},
\end{aligned}
\end{equation}
and are still widely used in observations and boundary layer schemes in numerical models to calculate the turbulent fluxes from the profiles of mean wind speed and temperature. Under the assumption that the mean shear and stress tensor are aligned ($u_*^2 \sim -\overline{u'w'}$ when streamline coordinates are used), the eddy diffusivities can be related to flux-gradient relations (Equation~\ref{eq:grad}) in the following way:

\begin{equation}
\label{eq:eddy}
\begin{aligned}
    K_m &= \dfrac{u_*kz}{\phi_m},\\
    K_h &= \dfrac{u_*kz}{\phi_h}.
\end{aligned}
\end{equation}
Given the strong dependence of $\phi_M$ and $\phi_H$ on anisotropy, it is instructive to explore the dependence of eddy diffusivities on anisotropy as well. 

\begin{figure}
\centering
\includegraphics[width=0.9\textwidth]{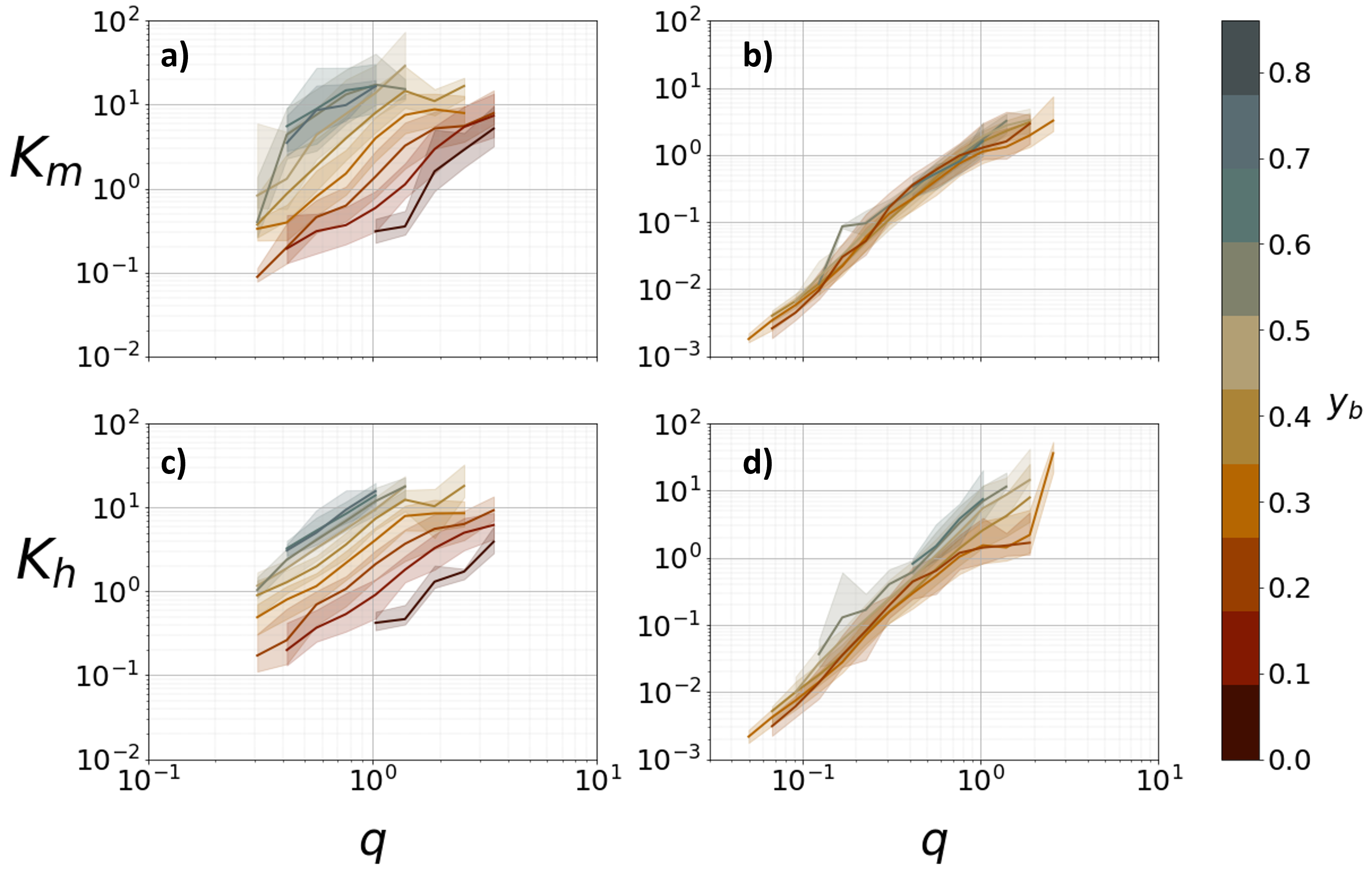}
\caption{Eddy diffusivities for momentum (a,c) and heat (b,d) under unstable (a,c) and stable (b,d) conditions, as defined in Eq.~\ref{eq:eddy}. Data are binned according the $y_b$, while medians of the logarithmically binned $\zeta$ are shown in full line, and inter-quartile range in shading.}
\label{fig:eddy}
\end{figure}

First order closure boundary layer schemes use the TKE (determined either diagnostically, prognostically or iteratively) together with a length scale, to parameterize the eddy diffusivities and calculate the turbulent fluxes from the mean gradients \citep[see][for a comparison of state of the art schemes]{maroneze2021two}. We therefore investigate the dependence of $K_m$ and $K_h$ on $q = \sqrt{TKE}$ and $y_b$ (Fig. \ref{fig:eddy}).
The dependence of both $K_h$ and $K_m$ on $y_b$ is stronger under unstable conditions, where anisotropy is as important as TKE, accounting for more than an order of magnitude of variability in $K$. In the stable regime on the other hand, only $K_h$ shows significant dependence on $y_b$ under conditions of well developed turbulence (higher $q$). Independently from the stability regime and the magnitude of the effect of anisotropy on eddy diffusivities, it is generally true that both $K_m$ and $K_h$ for isotropic turbulence are higher than for anisotropic turbulence. 
Figure \ref{fig:eddy} shows that the dependence of the eddy diffusivities on $q$ is consistent when turbulence anisotropy is taken into account. While in most numerical schemes $K_h$ and $K_m$ are parameterized as functions of $q \text{ or } q^{-1}$ the available data suggests a power law closer to $q^2$, i.e. a linear dependence on TKE, and an increasing function of $y_b$ especially relevant in unstable conditions. 

The analysis of the eddy diffusivities highlights an important feature of turbulence in the surface layer: isotropic turbulence is generally more efficient at transporting momentum and heat down-gradient than its anisotropic counterpart for the same stability conditions ($\zeta$, not shown) or turbulent energy ($TKE$). In essence, isotropic turbulence can either produce more momentum (buoyancy) flux for the same gradient or produce the same momentum (buoyancy) flux with lower gradients than anisotropic turbulence. Still the results of section~\ref{sec:components} should be kept in mind, showing that isotropic turbulence generally exists in conditions of lower gradients and fluxes.

We saw in Sections \ref{sec:intro} and \ref{sec:results} that the turbulent Prandtl number $Pr_t$ plays a crucial role in explaining the power law of the scaling relation in unstable stratification. We thus explore the dependence of $Pr_t$ on anisotropy. $Pr_t$ represents the ratio of diffusivities of momentum and heat, and is defined as:
\begin{equation}
\label{eq:prandtl}
Pr_t = \frac{K_m}{K_h} =  \frac{\phi_h}{\phi_m}.
\end{equation}

\begin{figure}
\centering
\includegraphics[width=0.9\textwidth]{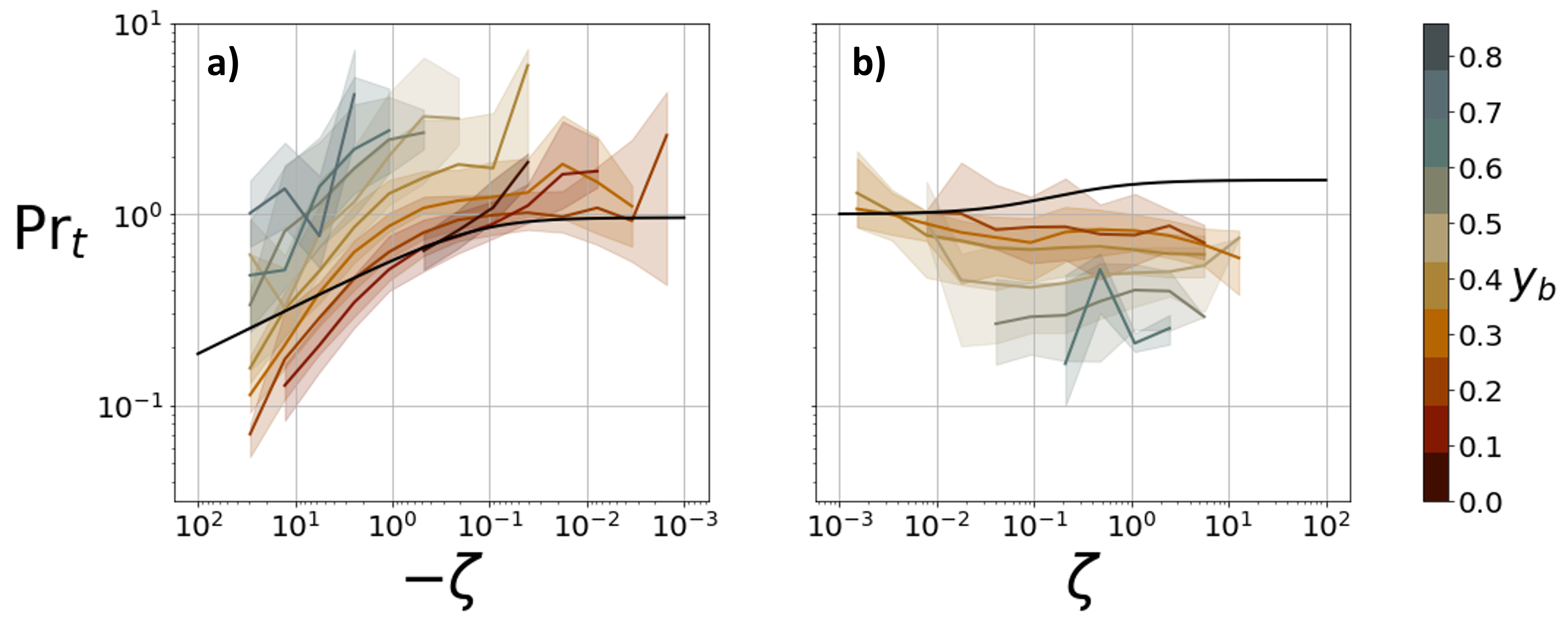}
\caption{The turbulent Prandtl number under unstable (a) and stable (b) conditions, as defined in Eq.~\ref{eq:prandtl}. Data are binned according to $y_b$, while medians of the logarithmically binned $\zeta$ are shown in full line, and inter-quartile range in shading. The black solid line represents the HO96 model.}
\label{fig:prandtl}
\end{figure}

Figure~\ref{fig:prandtl} shows the dependence of $Pr_t$ on stability and anisotropy in the ensemble data. The data show substantial scatter, especially in near neutral conditions as reported by \cite{li2019turbulent} and follow the theoretical curve from HÖ96 poorly. Still, as expected, in near-neutral unstable conditions turbulence mixes momentum better than heat, while the opposite happens in the very unstable regime ($\zeta<-1$) and in the stable regime. None the less, the data show that the turbulent Prandtl number has a strong dependence on turbulence anisotropy, which accounts for up to an order of magnitude of variation in the convective regime.
The results clearly show that $Pr_t$ approaches zero when $\zeta \to -\infty$, which contradicts the fundamental assumption of the O'KEYPS equation of a constant asymptotic limit. This is fundamentally linked to the power law of $\phi_M$ in the convective regime: data clearly shows that $\phi_M$ is following a $-\zeta^{1/3}$ power law as supported by scaling arguments (cf.~Section~\ref{sec:results}), while $\phi_H$ follows a well agreed upon $-\zeta^{-1/3}$ power law, leading to a vanishing denominator in $Pr_t$. 

Finally, it is interesting to notice the opposite role of isotropic structures in the stable and unstable boundary layer: in unstable conditions, given the same $\zeta$, $Pr_t$ for isotropic turbulence, i.e. the ratio of the efficiencies of mixing momentum versus heat, is much greater compared to anisotropic turbulence. Alternatively, in the stable regime the opposite happens: isotropic turbulence exhibits a lower $Pr_t$ than anisotropic turbulence.

\section{Conclusions}
\label{sec:conclusions}
This work presents a generalization of flux-gradient relations of the Monin-Obukhov similarity theory outside of its original range of applicability. The new scaling relations for the non-dimensional wind shear $\phi_M$ and temperature gradient $\phi_H$ were developed by including the degree of anisotropy as an additional parameter to the local similarity scaling framework, following the work of \cite{stiperski2023generalizing}. In the unstable regime ($\zeta<0$) the dependence on anisotropy was added to the model of \cite{kader90}, while in the stable regime ($\zeta>0$) a multivariate linear model was used. The new scaling relations show substantial improvement over classic MOST formulations and were shown to explain majority of the observed scatter in the flux-gradient relations. The improvement was especially large in highly convective conditions, and for scaled temperature gradient $\phi_H$ in both the unstable and stable regime. This clear dependence of $\phi_H$ on the degree of anisotropy explains the commonly observed large scatter in $\phi_H$ and therefore also the large uncertainty of heat fluxes derived from it. Finally, the inclusion of the componentality of turbulence through the first invariant of the barycentric map $x_b$ did not lead to significant improvement in stable stratification, and was therefore not considered further. 

The novel scaling additionally allowed to unravel some important features of the atmospheric boundary layer. In particular, they were able to resolve the long standing debate on the correct power law of $\phi_M$ in the free convective limit ($\zeta<-1$). Our results show that in contrast to analytical models based on the O'KEYPS equation, the correct power law is given by $\phi_M\sim-\zeta^{1/3}$ as predicted by \cite{kader90}. Additionally, the analysis showed that this rise in $\phi_M$ in highly convective conditions stems from the flattening of the wind shear for $\zeta \to -\infty$, while $u_*$ approaches zero. 
In unstable regime isotropic turbulence was also shown to have a lower wind shear and temperature gradient, and lower heat flux than anisotropic turbulence. The analysis of eddy diffusivities for momentum and heat showed a clear dependence on anisotropy and higher values for more isotropic turbulence, meaning that isotropic turbulence is in general more efficient at creating mixing while existing in conditions of smaller gradients and smaller fluxes. On the other hand, in the stable regime, isotropic turbulence occurred under similar wind speed conditions as anisotropic turbulence, with similar wind shear and slightly higher friction velocity in very stable stratification and consistently lower temperature gradients. The buoyancy flux $\overline{w'\theta'}$ achieved its maximum values \citep[cf.,][]{grachev05} at different stability according to turbulence anisotropy: the maximum in $\overline{w'\theta'}$ was at $\zeta \approx 0.05$ in very anisotropic turbulence while for quasi-isotropic turbulence it occurred at $\zeta \approx 0.5$. In this regime the dependence on anisotropy of the eddy diffusivities is much weaker than in the unstable regime, with a strong dependence on TKE.
The turbulent Prandtl number was shown to go to zero in the free convective regime, in contrast with the assumption of the O'KEYPS equation, and to have different behaviors in the two regimes: in unstable conditions isotropic turbulence is associated with a higher Prandtl number while the opposite happens in the unstable regime.

The new scaling relations and the conclusions derived from them show robust and promising pathways for the study of boundary layer turbulence. Still, a number of limitations inherent to the MOST approach remain, especially prominent in stable stratification. The commonly applied quality criteria used in all scaling studies, including the present one, do not allow a thorough study of the full range of conditions encountered in the atmospheric boundary layer. For example, stationary data are only observed in a small fraction of cases in stable stratification. In addition, counter-gradient fluxes, which tend to happen during nighttime and produce positive buoyancy flux with very anisotropic, mostly one-component, turbulence, are excluded from the present analysis as they would be erroneously classified as daytime convective turbulence. There is currently no approach that would allow including them into the MOST framework. Finally, in complex terrain, katabatic winds with a low level jet profile form on slopes during nighttime. The measurements taken above the jet maximum produce a negative wind shear that is filtered out by our quality criteria. Further analysis is therefore needed to explore if a scaling framework for flow above the low level jet maximum could be developed.

In this work we have proposed a generalized flux-gradient formulation of MOST scaling relations, as well as a revised formulation of the first order eddy-diffusivity turbulence closure models based on the inclusion of turbulence anisotropy. This approach has the important advantage of overcoming the traditional canonical limitations of MOST formulation, where the assumptions of horizontal homogeneity are embedded. 
The new generalized scaling relations represent a leap forward in the way that surface boundary conditions are represented in ESMs at all grid scales, delivering a more accurate representation of the momentum, energy, and mass exchanges at the land-atmosphere interfaces, at the expense of an additional unknown, that is turbulence anisotropy.
From a practical standpoint, the question that remains, is how to determine turbulence anisotropy on-the-fly in numerical simulations, and how will this be affected by the traditional anisotropy of the numerical grids employed in ESMs, and their resolution. Turbulence anisotropy is scale-dependent, and thus its contribution to the overall revised flux-gradient relations will be affected by those factors. Nonetheless, these challenges cannot be studied experimentally through traditional single-tower measurements, but instead require large arrays of turbulence measurements, limiting the experimental analysis. 
At present, to investigate these remaining challenges we envision different approaches consisting in the use of high resolution numerical simulations, machine learning techniques, Lagrangian averaging schemes, and the use of algebraic stress models, while considering the dependence of turbulence anisotropy on terrain, stability, spectral properties, surface thermal and roughness heterogeneity and many more complex terrain features.

Coupling the extended MOST with reliable complex terrain turbulence anisotropy will fulfill a 21st century milestone in the boundary layer community: develop a scaling framework that works in complex terrain.

\printendnotes

\bibliography{Biblio.bib}

\graphicalabstract{Figures/Mosso_Fig2.png}{Monin Obukhov similarity theory (MOST) is widely used in earth system models to parametrize turbulent exchanges in the atmospheric surface layer, however its original range of applicability is limited to flat and homogeneous terrain and a restrictive stability range. In this work we extend the original MOST to complex terrain by including information on turbulence anisotropy into the the flux-gradient scaling relations, showing substantial improvement, especially in conditions of unstable stratification. }

\end{document}